\documentclass[twocolumn,aps,prd,showpacs,showkeys,nofootinbib]{revtex4}
\usepackage{amsfonts,amssymb}

\newcommand{\Theorem}[2]{\medskip\noi {\bf #1. \ }{\sl #2}\medskip}

\def\nqq{\hspace*{-2em}}

\def\cm{\hspace*{1cm}}

\def\noi{\noindent}

\def\Jl#1#2{#1 {\bf #2},\ }

\def\ApJ#1 {\Jl{Astroph. J.}{#1}}
\def\CQG#1 {\Jl{Class. Quantum Grav.}{#1}}
\def\DAN#1 {\Jl{Dokl. AN SSSR}{#1}}
\def\GC#1 {\Jl{Grav. \& Cosmol.}{#1}}
\def\GRG#1 {\Jl{Gen. Rel. Grav.}{#1}}
\def\JETF#1 {\Jl{Zh. Eksp. Teor. Fiz.}{#1}}
\def\JETP#1 {\Jl{Sov. Phys. JETP}{#1}}
\def\JHEP#1 {\Jl{JHEP}{#1}}
\def\JMP#1 {\Jl{J. Math. Phys.}{#1}}
\def\NPB#1 {\Jl{Nucl. Phys.}{B\ #1}}
\def\NP#1 {\Jl{Nucl. Phys.}{#1}}
\def\PLA#1 {\Jl{Phys. Lett.}{#1A}}
\def\PLB#1 {\Jl{Phys. Lett.}{#1B}}
\def\PRD#1 {\Jl{Phys. Rev.}{D\ #1}}
\def\PRL#1 {\Jl{Phys. Rev. Lett.}{#1}}

\def\lal{&&\nqq {}}
\def\eq{Eq.\,}
\def\eqs{Eqs.\,}
\def\beq{\begin{equation}}
\def\eeq{\end{equation}}
\def\bear{\begin{eqnarray}}
\def\bearr{\begin{eqnarray} \lal}
\def\ear{\end{eqnarray}}
\def\earn{\nonumber \end{eqnarray}}
\def\nn{\nonumber\\ {}}

\def\tst{\textstyle}

\def\fract#1#2{{\tst\frac{#1}{#2}}}

\def\half{{\fract{1}{2}}}

\def\d{\d}

\def\diag{\mathop{\rm diag}\nolimits}

\def\const{{\rm const}}

\def\then{\ \Rightarrow\ }

\def\mn{_{\mu\nu}}
\def\mN{_{\mu}^{\nu}}
\def\MN{^{\mu\nu}}

\def\vac{{}_{\rm (vac)}}
\def\tot{{}_{\rm (tot)}}

\def\N{{\mathbb N}}

\def\ssph{static, spherically symmetric}
\def\bh{black hole}
\def\bhs{black holes}
\def\KS{Kan\-tow\-ski-Sachs}

\begin{document}

\title{Black holes can have curly hair}

\author{K.A. Bronnikov}
\affiliation
	{Center for Gravitation and Fundamental Metrology, VNIIMS, 46 Ozyornaya St., Moscow 119361, Russia;\\
        Institute of Gravitation and Cosmology,	PFUR, 6 Miklukho-Maklaya St., Moscow 117198, Russia}
	\email {kb20@yandex.ru}

\author{O.B. Zaslavskii}
\affiliation
{Astronomical Institute of Kharkov V.N. Karazin National University,
	35 Sumskaya St., Kharkov, 61022, Ukraine}
	\email {ozaslav@kharkov.ua}

\begin{abstract}
    We study equilibrium conditions between a \ssph\ \bh\ and classical
    matter in terms of the radial pressure to density ratio $p_r/\rho =
    w(u)$, where $u$ is the radial coordinate. It is shown that such an
    equilibrium is possible in two cases: (i) the well-known case $w\to -1$
    as $u\to u_h$ (the horizon), i.e., ``vacuum'' matter, for which
    $\rho(u_h)$ can be nonzero; (ii) $w \to -1/(1{+}2k)$ and $\rho \sim
    (u-u_h)^k$ as $u\to u_h$, where $k>0$ is a positive integer ($w=-1/3$
    in the generic case $k=1$). A non-interacting mixture of these two kinds
    of matter can also exist. The whole reasoning is local, hence the
    results do not depend on any global or asymptotic conditions. They mean,
    in particular, that a static black hole cannot live inside a star with
    nonnegative pressure and density. As an example, an exact solution for
    an isotropic fluid with $w = -1/3$ (that is, a fluid of disordered
    cosmic strings), with or without vacuum matter, is presented.
\end{abstract}


\maketitle

  In real astrophysical conditions, \bhs\ do not exist in empty space but
  are rather surrounded by some kind of matter which is either in
  equilibrium with the \bh\ or is falling on it. In other words, real \bhs\
  are ``dirty''. Meanwhile, the famous no-hair theorems (see, e.g.,
  \cite{fn,bek} and references therein) are not directly applicable to
  such situations of evident astrophysical interest. The main route in
  generalizing the possible black hole ``hair'' in such theorems consists in
  considering different (dilaton, gauge etc.) fields, whereas a much simpler
  but physically and astrophysically more relevant environment, namely,
  macroscopic matter with certain pressure and density, drops out from
  consideration.

  The aim of the present paper is to partly fill this gap and to prove some
  statements of this kind in the framework of general relativity. Strange as
  it may seem, to the best of our knowledge, they were not found (or at
  least explictly formulated) before.

  The conditions we will rely on are the horizon regularity, the Einstein
  equations and the conservation law for matter. For simplicity, we restrict
  ourselves to \ssph\ configurations. The manner of reasoning is close
  to that of Ref.\,\cite{ko}, where we have obtained some model-independent
  restrictions on the kinds of matter able to support regular cosmological
  Killing horizons in \KS\ geometries.

  We begin with writing the general \ssph\ metric in the form
\beq                                                         \label{ds}
   ds^2 = A(u) dt^2 - \frac{du^2}{A(u)}
   			     - r^2(u) (d \theta^2 + \sin^2 \theta d\phi^2),
\eeq
  where we have chosen the quasiglobal radial coordinate, corresponding to
  the ``gauge'' condition $g_{00}g_{11} = -1$. It has the following
  important properties \cite{vac1,cold}: it always takes a finite value $u =
  u_h$ at a Killing horizon where $A(u) =0$;\footnote
       {In principle, $u$ can take an infinite value at a candidate horizon
        where $A\to 0$, but then, as one can check, the canonical parameter
        of the geodesics also tends to infinity, so that the space-time is
  	already geodesically complete and no continuation is required.
	Such cases, which can be termed ``remote horizons'', can be found,
	e.g., in some solutions of the Brans-Dicke theory \cite{cold}. We
	will not discuss them here.	}
  moreover, near a horizon, the increment $u - u_h$ is a multiple (with a
  nonzero constant factor) of the corresponding increments of manifestly
  well-behaved Kruskal-type null coordinates, used for analytic continuation
  of the metric across the horizon. Therefore, with this coordinate, the
  geometry can be considered jointly on both sides of a horizon in terms of
  a formally static metric (hence the name ``quasiglobal''). On the other
  hand, both $A(u)$ and $r(u)$ should be analytic (or smooth at least up to
  derivatives of a certain order $s\geq 2$) functions of $u$ at $u=u_h$.  A
  regular horizon corresponds to a regular zero of $A(u)$, i.e., $A (u) \sim
  (u-u_h)^n$, where $n \in \N$ is the order of the horizon. In the case of a
  \bh, the outermost zero of $A$ corresponds to the event horizon.

  Consider the Einstein equations\footnote
  	{We use the units $c= \hbar = G =1$.}
  $G\mN \equiv R\mN - \half \delta\mN R = -8 \pi T\mN$ for the metric
  (\ref{ds}), so that, due to the symmetry of the problem, the stress-energy
  tensor (SET) for an arbitrary kind of matter can be written as
\beq                                                      \label{SET}
     T\mN = \diag (\rho,\ -p_r,\ -p_{\bot}, -p_{\bot}),
\eeq
  where the density $\rho$, the radial pressure $p_r$ and the transverse
  pressure $p_{\bot}$ are functions of $u$.

  One of the Einstein equations reads
\beq                                                       \label{01}
        G_0^0 - G_1^1 \equiv 2 A\frac{r''}{r}  = - 8\pi (\rho + p_r),
\eeq
  where the prime stands for $d/du$. \eq (\ref{01}) leads to a regularity
  condition at $u = u_h$ in terms of $\rho$ and $p_r$:
\beq
	p_r(u_h) + \rho (u_h) = 0,    			\label{hor}
\eeq
  or, more precisely, $p_r + \rho$ must have a zero of at least the same
  order as $A(u)$ since, by regularity, $|r''| < \infty$.

  Regularity thus requires that, at a horizon, the null energy condition
  (NEC)
\beq
       T\mN \xi^{\mu} \xi_{\nu } \geq 0,\cm
       		\xi^{\mu} \xi_{\mu} = 0.   		  \label{null}
\eeq
  should be obeyed on the verge. Indeed, for the SET (\ref{SET}), the NEC
  leads to
\beq
	p_{r}+\rho \geq 0, \cm      p_{\bot }+\rho \geq 0.  \label{NEC2}
\eeq

  The condition (\ref{hor}) is often discussed in connection with
  back-reaction of quantum fields (see, e.g., \cite{vis96}). Then, there is
  no great sense to speak of an equation of state; we simply have three
  different quantities in (\ref{SET}), obtained from quantum mean values
  in a fixed background, as functions of the radial coordinate. In what
  follows, we discuss relations between them, mostly between $p_r$ and
  $\rho$, in quite a general form, and our reasoning is equally applicable
  to quantum mean values, classical fields and (which is astrophysically
  more relevant) usual matter with certain equations of state
  $p_r = p_r(\rho)$ and $p_{\bot}=p_{\bot}(\rho)$.

  We are interested in general properties of matter surrounding a black hole
  horizon. A question of great importance for astrophysics is which kind of
  matter is consistent with the existence of a horizon. Although the condition
  (\ref{hor}) excludes $p_r$ and $\rho$ being both positive, it can be
  satisfied if both quantities tend smoothly to zero as one approaches the
  horizon. This will be the subject of our study. Accordingly, mostly
  bearing in mind small $\rho$ and $p_r$, we will use the linear relation
\beq
	p_r = w\rho,  \cm w = \const.  			\label{w}
\eeq

  We will assume $\rho \geq 0$. If, in addition, the NEC is satisfied
  (so that the weak energy condition is satisified as well), we call the
  matter normal, otherwise it is said to be phantom. One can note that the
  NEC is often violated due to quantum effects \cite{vis96}; phantom matter
  is also used in many studies as possible dark energy responsible for the
  accelerated expansion of the Universe.

\medskip\noi
  {\bf Consequences of the conservation law.}
  The only nontrivial component of the conservation law $\nabla_\nu T\mN =0$
  can be written in the form (the prime denotes $d/du$)
\beq
	p'_r + \frac{2r'}{r} (p_r - p_{\bot})
		+ \frac{A'}{2A} (\rho + p_r) = 0.	  \label{cons}
\eeq
  Suppose the validity of \eq (\ref{w}), at least near the
  horizon. As to the transverse pressure, we only assume that (at least, in
  the limit $\rho\to 0$)
\beq
	|p_{\bot}|/\rho  <  \infty.                       \label{p_bot}
\eeq
  It is a very weak restriction: indeed, for comparison, the dominant energy
  condition would require $|p_{\bot}|/\rho \leq 1$.

  Then, near the horizon, the term with $r'$ in \eq (\ref{cons}) can be
  neglected as compared with the third one. In the leading approximation
  (i.e., retaining terms of the order $\rho/\Delta u$), we obtain, as
  $A \to 0$,
\beq
	\rho \sim  A^{-(w+1)/(2w)}, \cm   w \ne 0.      \label{rho_h}
\eeq
  The value $w=0$ (dust) is naturally excluded since
  non-interacting dust cannot be in equilibrium in a static gravitational
  field. For different $w$, it follows from \eq (\ref{rho_h}):

\medskip\noi
{\bf (i)}
  $w > 0$ (normal matter with $p_{r} > 0$) or $w < -1$ (phantom). The density
  diverges as $A\to 0$. Thus such matter cannot exist near a horizon.

\medskip\noi
{\bf (ii)}
  $-1 < w < 0$ (normal matter with $p_r < 0$), in this case, both $\rho$ and
  $p_r$ tend to zero at the horizon, and the condition (\ref{hor}) holds.

\medskip\noi
{\bf (iii)}
  $w = -1$: this special case requires more attention. First, (\ref{hor})
  now may hold with $p_r = -\rho \ne 0$, which corresponds to a ``vacuum
  fluid'' considered in the next paragraph. Second, if we still assume $\rho
  \to 0$ as $A \to 0$, one can check that such a solution to \eq (\ref{cons})
  near $u=u_h$ can exist, but it cannot conform to the regularity
  requirements connected with the Einstein equations to be discussed below.
  Indeed, let us, going ahead, take $\rho$ in the form (\ref{rho_k}) and
  also assume a Taylor expansion of $p_r(\rho)$ at small $\rho$,
\beq
	p_{r} = -\rho + b\rho^{2} + \ldots.                  \label{Tayl}
\eeq
  Substituting all this into \eq (\ref{cons}) and equating coefficients
  by equal powers of $\Delta u$, we immediately obtain $\rho_k=0$, which
  means that $\rho(u)$ cannot be represented by a Taylor series near $u=u_h$.
  It is a general result which does not depend on $k$ or on the behavior of
  the function $r(\rho)$. One can find an asymptotic solution to \eq
  (\ref{cons}) for small $A$ under the assumption (\ref{Tayl}), having the
  form $\rho \approx -2/(b \ln A)$, so that really $\rho \to 0$ as $A\to 0$;
  however, in accord with the above general result, this solution does not
  have the form (\ref{rho_k}) and thus should be rejected.

\medskip\noi
  {\bf Inclusion of a vacuum fluid.}
  \eqs (\ref{hor}) and (\ref{cons}) for matter under consideration do not
  change if we add a ``vacuum'' matter with the SET \cite{dym}
\beq                                                     \label{SET-v}
      T\mN\vac = \diag (\rho\vac, \rho_r\vac, -p_\bot\vac, -p_\bot\vac),
\eeq
  and if there is no interaction between $T\mN$ and $T\mN\vac$, that is,
  the conservation law holds for each of them separately:
  $\nabla_\nu T\mN =0$ and  $\nabla_\nu T\mN\vac =0$.

  The definitive property of vacuum matter is that the SET (\ref{SET-v})
  preserves its form under arbitrary radial boosts \cite{dym}. Examples of
  such matter are usual Maxwell radial electric and magnetic fields for
  which $p_\bot\vac = \rho\vac$, their analogs in nonlinear electrodynamics
  with Lagrangians of the form $L_e = L_e(F)$, $F\equiv F\mn F\MN$
  \cite{br-NED}, Yang-Mills fields with a similar structure of the SET, and
  clouds of radially directed cosmic strings \cite{letelier79}.
  Independently of a particular realization of such vacuum matter, a number
  of important properties follow from its algebraic structure, $T_0^0\vac =
  T_u^u\vac$ \cite{dym}; it can be used both for constructing globally
  regular black hole models \cite{br-NED,dym} and for describing dark energy
  \cite{dym,br-dym07}.

  By definition, vacuum matter does not contribute to the third term in
  (\ref{cons}). Therefore, the above conclusions are valid for matter
  with the SET (\ref{SET}) independently of whether or not there is a
  non-interacting admixture of a vacuum fluid with the SET (\ref{SET-v}).

\medskip\noi
{\bf Consequences of the Einstein equations.}
  There are two independent components of the Einstein equations for the
  metric (\ref{ds}), with the unknown functions $A(u)$ and $r(u)$. Assuming
  the total SET $T\mN\tot = T\mN + T\mN\vac$, we can choose such two
  components as \eq (\ref{01})
  and the equation containing only first-order derivatives,
\beq
      G^1_1 \equiv                                            \label{11a}
              \frac{1}{r^2}[-1 + A'rr' + Ar'{}^2] = -8\pi(\rho\vac - p_r).
\eeq

  Now, we require that the metric should be analytic, or at least
  sufficiently smooth, in terms of the quasiglobal coordinate $u$
  (whose distinguished nature is discussed above, after \eq  (\ref{ds}))
  and thus admit continuation through the horizon. Therefore, we can write
  the Taylor expansions in $\Delta u \equiv u-u_h$:
\bear                                                      \label{A,r_h}
     A(u) &=& A_n \Delta u^n [1+ o(1)],
\nn
     r(u) &=& r_h + r'_h \Delta u + \half r''_u \Delta u^2 + o(\Delta u^2),
\ear
  with finite constants $A_n >0$, $r_h >0$, $r'_h$ and $r''_h$. Recall that
  $n \in \N$ is the order of the horizon.

  The l.h.s. of \eqs (\ref{01}) and (\ref{11a}) are also smooth at the
  horizon. The same then applies to the r.h.s., in other words, both $\rho$
  (hence $p_r$) and $\rho\vac$ are smooth and, in particular, since we
  are interested in configurations with $\rho\to 0$ as $u\to u_h$, we can
  write in the same limit
\beq                                                        \label{rho_k}
	\rho = \rho_k \Delta u^k [1+o(1)], \cm  k \in \N,
\eeq
  where $\rho_k = \const > 0$ and $k$ is the number of the first
  nonvanishing term of the Taylor series. Combining this with \eq
  (\ref{rho_h}), we obtain:
\beq                                                        \label{w_kn}
	k = - n\frac{w+1}{2w} \quad
			\then \quad  w = -\frac{n}{n + 2k},
\eeq
  a discrete set of values of $w = p_r/\rho\big|_{u=u_h}$.
  This whole set belongs to the range of interest, $-1 < w < 0$.

  Let us now substitute these quantities to the Einstein equations.
  \eq (\ref{01}) in the main approximation (with the largest terms kept
  on each side) gives
\beq                                                         \label{01h}
     2 A_n \frac{r''_h}{r_h}\Delta u^n = -8\pi (w+1)\rho_k \Delta u^k.
\eeq
  Evidently, finiteness of $r''_h$ leads to the requirement
\beq
	k \geq n,
\eeq
  where $k > n$ corresponds to $r''_h =0$.

  \eq (\ref{11a}), in turn, gives in its main approximation [where the
  first terms on both sides are simply rewritten while others are
  represented by their approximate expressions according to (\ref{A,r_h})
  and (\ref{rho_k})]
\beq                                                         \label{11h}
     -1 + nA_n r_h r'_h \Delta u^{n-1}
     		= -8\pi r_h^2 [\rho\vac(u_h) -w\rho_k \Delta u^k],
\eeq
  This relation leads to different results depending on the presence or
  absence of $\rho\vac$.

  Indeed, if $\rho\vac =0$ at the horizon, the r.h.s. of (\ref{11h}) is zero
  at $u=u_h$, the only way to satisfy the equation is to require $n = 1$,
  and then $A_1 r_h r'_h = 1$. The horizon is simple, Schwarzschild-like.
  Evidently, the generic case in the expansion (\ref{rho_k}) is $k=1$, and
  \eq (\ref{w_kn}) leads to $w= -1/3$ which, in case $p_r = p_\bot$,
  corresponds to a fluid of chaotically distributed cosmic strings (see
  \cite{-1/3} and references therein).

  In case $\rho\vac \ne 0$, any $n \geq 1$ is admissible; if $n > 1$, \eq
  (\ref{11h}) gives $\rho\vac(u_h) = 1/(8\pi r_h^2)$, while for $n = 1$ we
  have $\rho\vac(u_h) = (1 - A_1 r_h r'_h) /(8\pi r_h^2)$. Again, the
  generic case is certainly $n = k = 1$ and $w = -1/3$. One can also note
  that if matter is everywhere non-phantom, \eq (\ref{01}) leads to
  $r''<0$, and if, in addition, there is a spatial asymptotic (not
  necessarily flat) $r\to \infty$ as $u\to \infty$, then $r' >0$ in the
  whole space, and $r'_h >0$ in particular.

  We have actually proved the following theorem:

\Theorem {Theorem 1}
  {A \ssph\ \bh\ can be in equilibrium with a static matter distribution
   with the SET (\ref{SET}) only if near the event horizon ($u\to u_h$,
   where $u$ is the quasiglobal radial coordinate) either (i) $w \to -1$
   {\rm (matter in this case has the form of a ``vacuum fluid'')}
   or (ii) $w \to -1/(1+2k)$, where $w \equiv p_r/\rho$ and $k$ is a positive
   integer. In case (i), the horizon can be of any order $n$, and $\rho(u_h)$
   is nonzero. In case (ii), the horizon is simple, and $\rho \sim (u-u_h)^k$.}

  The generic case of such a non-vacuum hairy black hole is $k{=}1$,
  implying $w= -1/3$. In the isotropic case, $p_r = p_\bot$, it corresponds
  to a fluid of disordered cosmic strings \cite{-1/3}. Since such strings
  are, in general, arbitrarily curved and may be closed, one can express the
  meaning of the theorem by the words ``non-vacuum black holes can have
  curly hair''. Recall, however, that in general our $w$ characterizes the
  radial pressure, while the transverse one is only restricted by the
  condition (\ref{p_bot}).

  Other values of $k$ ($k = 2,\ 3$ etc.) represent special cases
  obtainable by fine tuning of the parameter $w$.

  In the presence of vacuum matter with the SET (\ref{SET-v}), the following
  theorem holds:

\Theorem {Theorem 2}
  {A \ssph\ \bh\ can be in equilibrium with a non-interacting mixture of
   static non-vacuum matter with the SET (\ref{SET}) and vacuum matter with
   the SET (\ref{SET-v}) only if, near the event horizon ($u\to u_h$),
   $w \equiv p_r/\rho \to -n/(n+2k)$, where $n \in \N$ is the order of the
   horizon, $n \leq k \in \N$, and $\rho \sim (u-u_h)^k$.}

  Thus a horizon of a static black hole can in general be surrounded by
  vacuum matter and matter with $w = -1/3$, which is true for any order of
  the horizon if $n=k$. (There also can be configurations with $k > n$ and
  fine-tuned equations of state where $w = -n/(n+2k) > -1/3$.) An
  arbitrarily small amount of other kinds of matter, normal or phantom,
  added to such a configuration, should break its static character by simply
  falling onto the horizon or maybe even by destroying the black hole. In
  other words, black holes may be hairy, or ``dirty'', but the possible
  kinds of hair are rather special in the near-horizon region:  normal (with
  $p_r\geq 0$) or phantom hair are completely excluded. In an equilibrium
  configuration, all ``dirt'' is washed away from the near-horizon region,
  leaving there only vacuumlike or modestly exotic, probably ``curly'' hair.

  In particular, a static \bh\ cannot live inside a star of normal matter
  with nonnegative pressure unless there is an accretion region around the
  horizon or a layer of ``string'' and/or vacuum matter.

  We did not discuss the behavior of $p_\bot$ and $p_\bot\vac$ (except for
  the restriction (\ref{p_bot})). In fact, these quantities are inessential
  for our reasoning but should be necessarily specified for finding complete
  solutions in particular models. Our inferences are quite general and hold
  for all kinds of hair: for instance, in all known examples of \bhs\ with
  scalar fields (see, e.g., \cite{pha1} and references therein), the SETs
  near the horizon must satisfy the above conditions, which may be directly
  checked.

  Also, our approach is relevant to semiclassical black holes in equilibrium
  with their Hawking radiation (the Hartle-Hawking state), whose SET
  essentially differs from that of a perfect fluid.  Since the density of
  quantum fields is, in general, nonzero at the horizon (see Sec.\,11 of the
  textbook \cite{fn} for details), the regularity condition (\ref{hor})
  tells us that such quantum radiation should behave near the horizon like
  a vacuum fluid.  Our results show that a black hole can
  be in equilibrium with a mixture of Hawking radiation and some kinds of
  classical matter with $-1 < w < 0$ (including the important case of a
  Pascal perfect fluid with $p_r = p_\bot$). Possible effects of this 
  circumstance for semiclassical black holes need a further study. Moreover, 
  large enough black holes, for which the Hawking radiation may  be neglected, 
  can be in equilibrium with classical matter alone, also including the 
  case of a perfect fluid.

  Our reasoning was entirely local, restricted to the neighborhood of the
  horizon, and the results, which involve the single parameter $w= p_r/\rho$,
  are in other respects model-independent. Meanwhile, a full analysis of
  specific systems would require the knowledge of the equation of state and
  conditions on the metric in the whole space (e.g., the asymptotic
  flatness condition). Such an analysis depends on the model in an essential
  way and is beyond the scope of this paper. One can add that the equations
  of state well-behaved near the horizon are often incompatible with
  reasonable conditions at infinity (see the example below); it simply means
  that such matter does not extend to infinity and can only occupy a finite
  region around the horizon.

  It would be of interest to generalize our results to nonspherical and
  rotating distributions of matter.

\medskip\noi
  {\bf Example.} In conclusion, let us present an exact solution for a
  system of utmost interest described by the above theorems. Consider a
  region of space with a non-interacting mixture of a vacuum fluid specified
  by $8\pi\rho\vac = \Lambda (u)$ [$p_\bot\vac$ is then found from the
  conservation law for the SET (\ref{SET-v})] and an isotropic fluid of
  cosmic strings, such that $p_r = p_{\bot} = -\rho/3$. Then \eq (\ref{cons})
  leads to $\rho = \rho_0 A$, \eq (\ref{01}) takes the form
  $r'' + \alpha^2 r =0$, and, without loss of generality, we have
\beq
	r(u) = r_0 \sin \alpha u,                            \label{r(u)}
\eeq
  where $\rho_0 > 0,\ r_0 > 0$ are arbitrary constants and
  $\alpha = (8\pi\rho_0/3)^{1/2}$. The remaining unknown function $A(u)$ can
  be found from \eq (\ref{11a}), which turns out to be linear,
\beq                                                         \label{11sol}
     	A'rr' + A(r'^2 + \alpha^2 r^2) = 1 - \Lambda(u) r^2,
\eeq
  hence easily integrable by quadratures for an arbitrary dependence
  $\Lambda(u)$ (or $\Lambda(r)$, as was used, e.g., in \cite{dym,br-dym07}).
  (Let us stress that our solution is different from that
  in Ref.\,\cite{letelier79}, obtained for a cloud of unidirectional strings.)
  In particular, for $\Lambda = \const$ we find
\beq                                                     \label{A(u)}
     A(u) = \frac{1}{\alpha^2 r_0^2}\biggl[
	       1- C \cot \alpha u
	     	    - \Lambda r_0^2(1 - \alpha u \cot \alpha u) \biggr],
\eeq
  with $C = \const$. A horizon corresponds to $A = 0$, e.g., in case
  $\Lambda = 0$ we obtain $u_h = (1/\alpha)\arctan C$; the horizon is
  simple: one can verify that $A'(u_h) \ne 0$. As follows from (\ref{r(u)}),
  this solution has no large $r$ asymptotic, but it can be incorporated in
  an asymptotically flat model by matching it at some $u > u_h$ to some
  intermediate layer (e.g., described by an analytic solution like the one
  for an incompressible fluid) admitting zero pressure at some surface, at
  which it can be further matched to the Schwarzschild solution.

  It seems instructive to trace the limiting transition from \eqs
  (\ref{r(u)}), (\ref{A(u)}) to the vacuum Schwarzschild--(anti) de Sitter
  metric. As the matter density vanishes, $\rho \to 0$, so that $\alpha\to
  0$, it is convenient to write $r_0 = \beta/\alpha$, $C = 2m\alpha/\beta$,
  where $\beta$ and $m$ are new constants. Then, after simple calculations,
  we obtain in the limit $\alpha\to 0$
\[
   r = \beta u, \qquad
  A(u)=\frac{1}{\beta^2}\biggl(1-\frac{2m}{u\beta}
  				-\frac{\Lambda}{3} \beta^2 u^2\biggr).
\]
  Rescaling  $t\mapsto \tilde{t}= t/\beta$ and using the coordinate $r$,
  we obtain the metric (1) with $A(r) =1 - 2m/r- \Lambda r^2/3$, as required.

\medskip
  One of the authors (K.B.) acknowledges partial financial support from
  Russian Basic Research Foundation Project 07-02-13614-ofi\_ts.

\end{document}